# AFM-based low frequency viscoelastic characterization of wood pulp fibers at different relative humidity

C. Czibula[1,2], T. Seidlhofer[2,3], C. Ganser[4], U. Hirn[2,3,*], C. Teichert[1,2]

[1]*Institute of Physics, Montanuniversitaet Leoben, Franz Josef Str. 18, 8700 Leoben, Austria*

[2]*Christian Doppler Laboratory for Fiber Swelling and Paper Performance, Graz University of Technology, Inffeldgasse 23, 8010 Graz, Austria*

[3]*Institute of Bioproducts and Paper Technology, Graz University of Technology, Inffeldgasse 23, 8010 Graz, Austria*

[4]*Exploratory Research Center on Life and Living Systems, National Institutes of Natural Sciences, Okazaki, Japan*

*\* corresponding author*

**Abstract**

The viscoelastic behavior of wood pulp fibers plays a fundamental role in the performance of paper and paper products. Wood pulp fibers are hierarchical composites consisting of different cell wall layers and have anisotropic properties. Since accessing the individual fibers is challenging, no measurement technique has been able to characterize the viscoelastic properties in both – the longitudinal and transverse – fiber direction yet. Here, an atomic force microscopy (AFM)-based method is applied to investigate the viscoelastic properties of wood pulp fibers at varying relative humidity (RH) in both fiber direction. Experimental creep tests have been performed to investigate the material's low frequency regime response. A spring-dashpot model has been applied to characterize the viscoelastic behavior. The results indicate surprisingly small differences of the properties between both fiber directions. Transferring the results into a spectral representation explains an opposing trend of the viscosity that is connected to the long-term behavior.

**Keywords**

*Atomic force microscopy, wood pulp fiber, viscoelastic creep, humidity dependence*

**Introduction**

Wood pulp fibers are a renewable resource and are mainly used in non-woven fabrics like paper, board, or sanitary tissues. Due to the production principle, these products have a composite layer-like structure with distinct fiber orientation, they also frequently contain mineral fillers. The viscoelastic properties of paper are important for applications and processing. For packaging applications, the failure over time plays an important role whereas for converting processes, the delayed deformation is critical. Even though viscoelastic properties of paper are well known [1], the characteristics of this behavior on the fiber level are not well understood. Consequently, the quantitative investigation of single fiber viscoelasticity will bring insight into this matter which is crucial e. g. for the development of mechanical models of nonwovens on the network level [2–4]. Furthermore, there are numerous research activities to use lignocellulosic fibers in composite materials [5–7] to eventually replace conventional reinforcements. Since in many technical applications the damping and relaxation behavior of the material is of high relevance, detailed knowledge on the viscoelastic properties of single fibers contributes to a better understanding and design of such reinforced composites. A further challenge are the hygroscopic properties of natural fibers [8]. Here, moisture absorption not only has an impact on the mechanical properties of the reinforcing fibers, but it also influences the adhesion between the fibers and the matrix [9]. The characterization of the interface between fibers and matrix is in general a challenging task [10].

Wood pulp fibers are industrially processed wood fibers. These fibers are a complex hierarchical composite and have highly anisotropic material properties. As illustrated later, the cell wall of an individual wood fiber is composed of cellulose microfibrils that are surrounded by a matrix of amorphous material (hemicellulose and lignin). The cell wall consists of different layers – primary (P) layer, secondary (S1, S2, and S3) layers – which differ in thickness, chemical composition, and cellulose microfibril alignment. The alignment of the microfibrils can be characterized by the microfibril angle (MFA) $\theta$. Furthermore, a hollow space called lumen (L) is located in the center of the fiber. The S2 layer constitutes about 80–95 % of the fiber mass and thus dominates the mechanical properties of the fiber. In this layer also the cellulose microfibrils are highly aligned which is the reason for the anisotropic mechanical behavior of the fiber. During the pulping process, the wood fiber undergoes several structural changes. First, the P layer usually gets removed during papermaking due to its high lignin content and random fibril alignment. Second, the paper production process leads to a collapse of the lumen [1,11,12]. Additionally, on a microstructural level, fiber porosity increases due to the removal of lignin [13].

Linear viscoelastic material models can be physically interpreted as a combination of linear springs and dashpots. The springs describe elastic behavior whereas the dashpots represent the viscous response. These elements can be combined arbitrarily and with either a so-called Generalized Maxwell (GM) model or a so-called Generalized Kelvin-Voigt model all possible behaviors can be reproduced for a linear viscoelastic solid material [14]. Since the GM model utilizes moduli as parameters rather than compliances, it is preferably used in this work. Viscoelastic behavior can be either tested with creep (constant stress applied), stress relaxation (constant strain applied), or sinusoidal excitation. Whereas dynamic measurements are mostly used for high-frequency response, the static creep or stress relaxation experimental routines investigate the viscoelastic behavior on an intermediate to long-term time scale.

For wood fibers, several investigations of the viscoelastic properties have been obtained. Green wood has been characterized in transverse direction [15], and the viscoelastic behavior of pine specimens has been described along the grain [16]. Furthermore, viscoelastic creep has been studied for beech wood under tensile and compression loading [17]. In general, it was found that compression wood tissue – exhibiting a higher MFA – compared to normal wood tissue shows pronounced viscoelastic relaxation [18–20]. Furthermore, for wood, the influence of hemicelluloses on the viscoelastic behavior has been studied [21], revealing that natural wood containing hemicelluloses exhibits a more pronounced viscoelastic behavior. In comparison, literature on the viscoelastic behavior of single wood pulp fibers is scarce although it is common knowledge that it is affecting the properties of paper products.

To access the microstructure of the fiber on the nanoscale, atomic force microscopy (AFM) [22] is a versatile technique. It does not only provide topographical information as obtained with morphological studies of the fiber surface [23], but also enables the implementation of more complex experiments, i. e. the measurement of the joint strength of single fiber-fiber bonds [24]. Since the AFM probe is an extremely sensitive force sensor, several AFM-based methods have been developed for the micromechanical characterization of wood pulp fibers [25–29]. Here, AFM-based nanoindentation (AFM-NI) experiments [28,30] have been applied to characterize the mechanical properties of pulp fibers under controlled humidity. It was found that the mechanical properties – reduced modulus and hardness – are decreasing with increasing RH level, in water the decrease is highest. Recently, a comprehensive AFM-based characterization of the viscoelastic properties in transverse direction at different RH and in water has been carried out [31]. The evaluation of the experimental data with a GM model resulted in a decreasing trend of the elastic and viscous parameters with increasing RH and a very pronounced drop of the values in water.

Most of the methods mentioned are bearing the limit that usually only the mechanical properties in one direction can be obtained. With conventional nanoindentation (NI), it was accomplished to measure the elastic stiffness tensor of wood fibers by using an approach based on anisotropic indentation theory on the micron scale [32–34]. However, the viscoelastic properties of wood pulp fibers in more than one direction at different RH levels have not been studied before.

In this work, the experimental protocol is force controlled and, therefore, the testing procedure has similarities to conventional creep test, but needs to be treated differently in the parameter evaluation. Here, a comprehensive AFM-NI creep study of the viscoelastic properties, at a low frequency regime (0.004–1 Hz), is presented for wood pulp fibers at different RH in longitudinal and transverse fiber direction. For that purpose, the data of the S1 layer in transverse direction obtained in [31] has been re-evaluated with a different GM model – a Generalized Maxwell model of order 3 (GM3) – to compare it to the new results that have been obtained on the S2 layer in longitudinal direction. This is the first time a viscoelastic characterization of a wood-based material has been obtained in longitudinal and transverse direction at different relative humidity levels measured with the same technique.

**Materials & Methods**

*-Wood pulp fiber samples*

In this work, two sets of samples have been tested. First, single pulp fibers (Mondi, Frantschach) have been investigated in transverse direction. The pulp fibers were industrial, unbleached, and unrefined softwood (spruce and pine) pulp with a kappa number κ = 42. The kappa number indicates the residual lignin content of the pulp which is for this pulp about 6 %. Different results of measurements on these fibers have been published before and a full description of the sample preparation and methodology can be found in [31]. To investigate the longitudinal direction, microtome cuts of paper produced from the same industrial pulp have been prepared. First, paper is embedded in a hydrogel-like material called glycol methacrylate (GMA), and then the paper is cut by a diamond knife to a slice thickness of about 7 µm [35,36]. For AFM measurements, these 7 µm thick microtome cut slices of paper are fixed onto a steel sample holder with nail polish – analogous to [28].

*-AFM*

All AFM based nanoindentation measurements reported in this work have been acquired with an Asylum Research MFP-3D AFM. The instrument is equipped with a closed-loop planar x–y-scanner with a scanning range of 85 x 85 µm² and a z-range of about 15 µm. For the viscoelastic characterization, LRCH250 silicon probes (Team Nanotec, Germany) with a tip radius of 300 nm have been used. The spring constant of the cantilever is (290.2 ± 51.3) N/m and has been calibrated by performing the thermal sweep method [37] (values are presented as mean ± standard deviation calculated from 4 independent calibration measurements for two different cantilevers). The thermal Q factor is 778 ± 224, and the resonance frequency is (575 ± 2) kHz.

To investigate fiber samples in an environment with defined RH, the AFM is equipped with a closed fluid cell (Asylum Research, USA) which can be flushed by nitrogen in a controlled way. This RH recording setup has been successfully employed before, and a more detailed description can be found in [28,30]. The employed load schedule (see Figure 1a) has been thoroughly discussed for the transverse direction in [31] and has been also applied with small adaptions for the longitudinal direction. The only modification of the load schedule is illustrated in Figure 1a. An increased force of 10 µN instead of 5 µN has to be applied for the viscoelastic testing in longitudinal direction due to otherwise too low indentation depths. To keep the loading time constant, the load rate was increased to 6.4 µN/s from 3.2 µN/s.

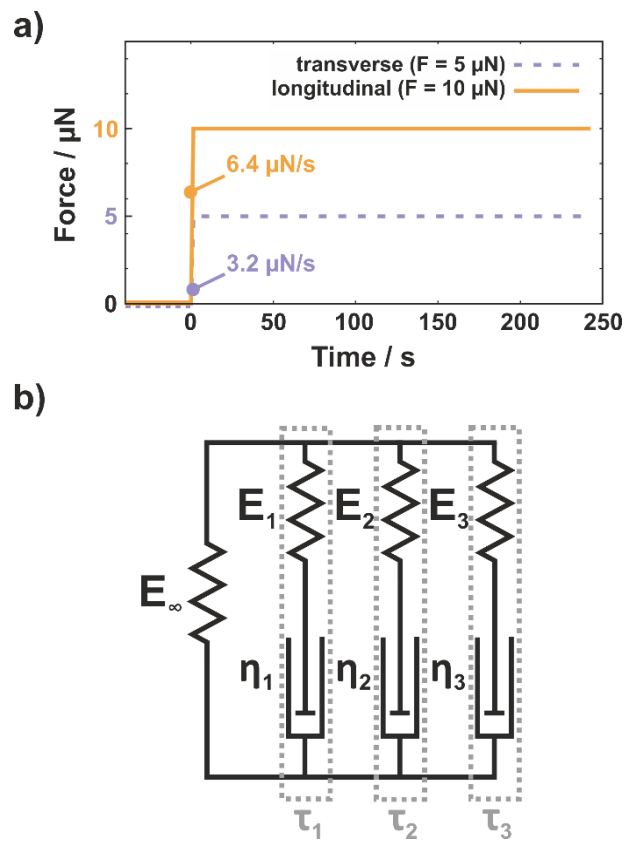

*Figure 1: (a) Load schedule of the viscoelastic AFM-NI experiment in transverse and longitudinal direction. (b) Illustration of the Generalized Maxwell model of order three (GM3) consisting of a single*

*spring $E_\infty$ parallel to three individual Maxwell elements. Each Maxwell element is characterized by an elastic parameter $E_i$, a viscous parameter $\eta_i$, and a characteristic relaxation time $\tau_i$.*

*-Viscoelasticity*

Viscoelasticity can be either characterized explicitly with integral type functionals (hereditary integrals) or implicitly with internal variables [38]. The functional representation of viscoelasticity [39,40]

$$\sigma = \int_0^t E(t-\tilde{t}) \frac{d\varepsilon(\tilde{t})}{d\tilde{t}} d\tilde{t} \tag{1}$$

reveals that a continuous function $E(t)$ – the time dependent elastic modulus – as a material property needs to be identified. The full investigation of $E(t)$ for $t \to \infty$ as well as the direct experimental investigation of $E(t)$ for short time scales are impossible. Therefore, limitations to a certain time domain have to be established. Additionally, it is convenient to reduce the storage amount for $E(t)$ by an appropriate parametrization. Here, commonly the Prony series [41]

$$E(t) = E_\infty + \sum_{i=1}^{n} E_i \exp\left(\frac{-t}{\tau_i}\right) \tag{2}$$

is utilized for this purpose. This series first introduces the equilibrium behavior $E_\infty$, which can be considered the linear elastic modulus of the material after infinite loading time. Furthermore, there are $n$ relaxation branches which are defined by magnitude $E_i$ and relaxation time $\tau_i$. Since the series approximates a continuous function, $\tau_i$ can be imagined as supporting points in the time domain. A simple way to imagine one relaxation branch $E_i \exp\left(\frac{-t}{\tau_i}\right)$ is that after the passage of the relaxation time $\tau_i$, the tension in the branch has dropped to $E_i \frac{1}{e}$.

It is advisable to logarithmically evenly distribute the $n$ relaxation modes over the relevant time scale avoiding numerical difficulties when the $E_i$ are identified out of experiments. Here, the logarithmic nature is a result of the exponential form of the Prony series.

By introducing the Prony series the functional representation can be turned into an internal variable representation. These internal variable representations lead to a set of $n$ ordinary differential equations for the evolution of the internal state space. This internal state can be physically interpreted by the generalized Maxwell model (GM), which is mathematically equivalent to the Prony series representation [42]. Therefore, the internal state of the GM is defined by the positions of the springs and dashpots. Here, the GM model of order three (GM3) – involving three relaxation modes – was found to fit the experimental data adequately and is illustrated in Figure 1b. It consists of a single spring $E_\infty$ parallel to three individual so-called Maxwell elements containing a spring in series with a dashpot. Each Maxwell element is characterized by an elastic parameter $E_i$ and a viscous parameter $\eta_i$ or by a characteristic relaxation time $\tau_i$ – which is the ratio of the viscous and elastic parameter. For further discussion of the results in this work, the GM3 model will be used.

To better visualize the results, a spectral representation of the viscoelastic parameters of the spring – dashpot model can be useful as well [43]. Here, for the calculation of the spectrum, the complex modulus $E^* = E' + iE''$ needs to be introduced [44]. It consists of the so-called storage modulus $E'$, which is a measure of the elastic response of the material, and the loss modulus $E''$, describing the viscous response of the material. For a GM model of different relaxation modes $n$, the following equations apply:

$$E'(\omega) = E_\infty + \sum_{i=1}^{n} E_i \frac{(\omega \tau_i)^2}{1+(\omega \tau_i)^2} ,  \qquad (3)$$

and

$$E''(\omega) = \sum_{i=1}^{n} E_i \frac{\omega \tau_i}{1+(\omega \tau_i)^2} , \qquad (4)$$

where $\omega$ represents the angular frequency. The loss tangent $\tan(\delta)$, which is associated with the ratio of the energies dissipated and stored and which provides a measure of the damping in the material can be defined as

$$\tan \delta(\omega) = \frac{E''}{E'}. \qquad (5)$$

**Results & Discussion**

In this work, the viscoelastic properties of wood pulp fibers have been investigated by AFM at varying RH in different directions relative to the long fiber axis as well as on the surface of different cell wall layers. In longitudinal direction, the S2 layer and in transverse direction, the S1 layer has been tested as is presented in Figure 2. Since the S2 layer is not directly accessible from the surface of the fiber, an approach to measure useful and reliable data needed to be developed. The approach chosen in this work is microtome-cutting thin slices of an embedded paper sheet consisting of single wood pulp fibers and investigating the fiber cross-sections. This way, the S2 layer in longitudinal direction can be investigated. This preparation routine is well established [35,36].

Measurements in both directions have been obtained at 25 %, 45 %, 60 %, and 75 % RH. Unfortunately, for the longitudinal direction, it was not possible to perform measurements at higher RH levels than 75 % RH since the embedding material GMA exhibits a high degree of swelling. Therefore, the increase in height of GMA makes it impossible for the AFM probe to access the fiber surface.

Figure 2b and c show representative 5 x 5 µm² AFM topography images of the investigated surfaces of a pulp fiber in longitudinal (Fig.2b) and in transverse direction (Fig. 2c) at 60 % RH. In Figure 2b, a fiber cross-section is visible. Due to the microtome-based cutting procedure, the surface of the S2 layer is rather smooth (root mean square (RMS) roughness about 15 nm) but exhibits some cracks. Furthermore, it is visible that the lumen is nearly completely collapsed. In comparison, the pulp fiber in transverse direction (as presented in Figure 2c) has a very rough surface which is dominated by wrinkle-like structures that are induced by the drying process of the fibers (RMS roughness about 150 nm).

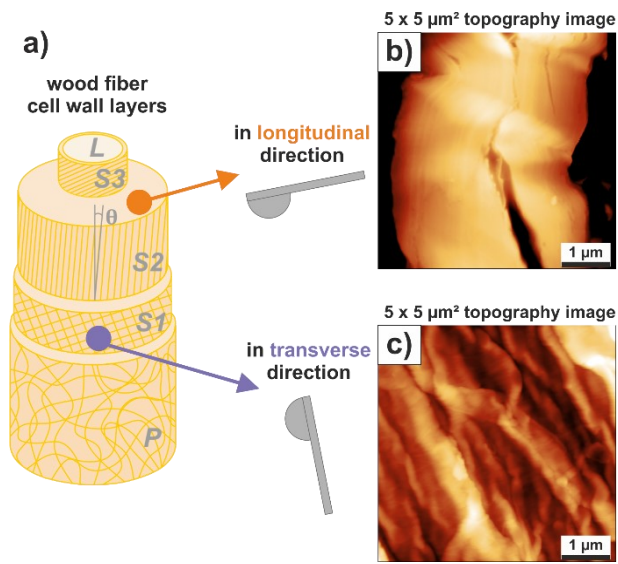

*Figure 2: (a) Illustration of the cell wall layers of a wood pulp fiber. It consists of a primary (P) and three secondary (S1, S2, S3) cell wall layers and a lumen (L). The microfibrillar angle (MFA) ϑ indicates the angle between the long fiber axis and the microfibril orientation. The dots indicate the measured layers with the corresponding directions indicated by the orientation of the AFM probe. In (b), a 5 x 5 µm² AFM topography image of a cross-sectional cut of a pulp fiber to access the S2 layer is presented. In (c), a 5 x 5 µm² topography image of the surface of a pulp fiber in transverse direction is presented. The z-scale is 600 nm for both AFM images.*

Since the experimental procedure for the measurements in transverse direction has been thoroughly described and discussed already [31], here, the focus will be on the longitudinal direction. In Figure 3, AFM topography images of the surface of the microtome cuts of fiber cross-sections are presented. To start a viscoelastic AFM experiment, an appropriate fiber surface needs to be located. Therefore, large overview AFM scans, as illustrated in Figure 3a, are performed. After scanning an area of up to 50 x 50 µm², a fiber cross-section can be selected, and the surface of this fiber is scanned again before the viscoelastic measurement routine (Figure 3b). Taking a closer look at Figure 3a and b, reveals that the fibers are lower compared to the embedding. It is possible to clearly distinguish between the surfaces. Furthermore, since it is known that the S2 layer is the thickest layer and the S1 and S3 layer are only several hundred nm in thickness [45], it is easy to determine the S2 region for measurements as indicated in Figure 3b with the dashed white line. It should be noted that for the viscoelastic data evaluation of each measured region (corresponding to a 5 x 5 µm² window like in Fig. 3b, c) an average curve of the individual measurement points (9 measurement points in Fig. 3c) has been calculated. The reason was to reduce the influence of thermal drift and signal noise, especially at lower RH, as described in more detail in [46]. At low RH, the fiber surface is so stiff that the AFM probe only penetrates a few nanometers. Therefore, no measurements in longitudinal direction have been obtained below 25 % RH. In general, the maximum force for the longitudinal direction had to be increased from 5 µN (employed for the transverse direction) to 10 µN to obtain a sufficiently high indentation depth (> 10 nm). Further details about measurement difficulties for AFM-NI experiments on wood pulp fibers are provided in [31].

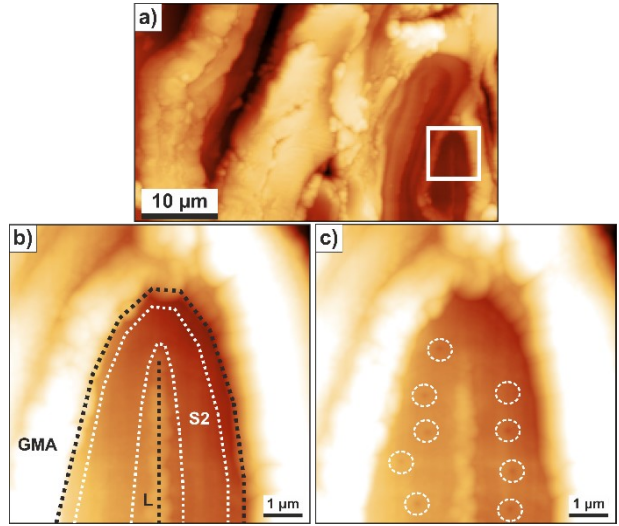

*Figure 3: Topography of the surface of microtome cuts to access the S2 layer in longitudinal direction. (a) 50 x 30 µm² AFM topography overview image (z-scale: 4.6 µm). The white square illustrates the zoom-in region in (b, c). In (b), an 8 x 8 µm² AFM topography image (z-scale: 1.8 µm) of part of a fiber cross-section is presented. The outer black dashed line indicates the border of the fiber surface to the embedding material GMA, whereas the single dashed black curve in the middle marks the lumen L. The white dashed lines indicate the region of the S2 layer where measurements are obtained. In (c), the same topography as in (b) has been re-measured after the viscoelastic experiment and all the single indents are highlighted by dashed white circles.*

A Raman spectra analysis has been performed to investigate whether GMA is penetrating the fiber cell wall. As presented in the spectra presented in Figure A1 in the Electronic Supplementary Information (ESI), the results show that GMA does not penetrate the fiber cell wall. Therefore, it is assumed that the embedding of the fibers in GMA to produce microtome slices has a negligible influence on the mechanical properties of the fibers during the AFM-NI experiments. Furthermore, the mechanical properties of GMA were also investigated by AFM-NI using the pyramidal probe. For GMA, the values of $M$ and $H$ are always lower than for the wood pulp fibers in both directions, and the material also exhibits a dependence on RH (Figure A2 in ESI).

In Fig. 4, averaged experimental creep curves from all experiments for all RH values in longitudinal and transverse directions are presented. It should be emphasized that the applied forces differ in both directions. Whereas in longitudinal direction, a force of 10 µN was applied to obtain sufficiently high indentation depths, in transverse direction, only 5 µN has been applied. In both directions, similar indentation depths are achieved. Furthermore, as expected, the indentation depth and the initial slope of the experimental curves are increasing with increasing RH. The creep curves in longitudinal direction (Figure 4a) between 25 % RH – 60 % RH are quite similar with an indentation depth of about 20 nm and exhibit a low slope. Only at 75 % RH, the creep curve has a pronounced slope and reaches an indentation depth of about 40 nm. In Figure 4b, the creep curves in transverse direction are presented. The creep curve at 25 % RH exhibits an indentation depth lower than 20 nm, whereas the curves for 45 % RH and 60 % RH are nearly identical and have a more pronounced slope. Comparing both directions, the curves between 25 % and 60 % RH show a higher slope compared to curves in the longitudinal direction. However, for both directions, the curves at 45 % RH and 60 % RH are nearly overlapping, indicating similar creep behavior. Furthermore, also the creep curves at 75 % RH appear quite similar for both directions.

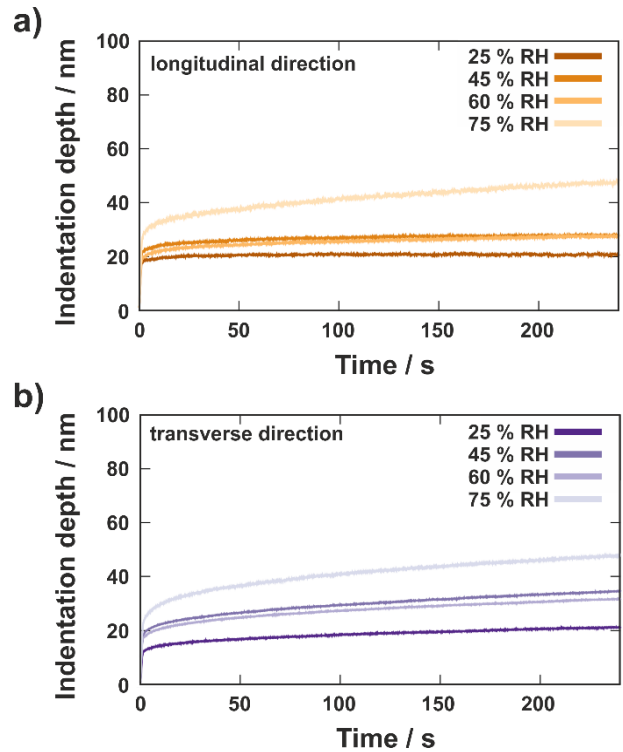

*Figure 4: Averaged experimental creep curves for the time interval of 240 seconds at 25 % RH, 45 % RH, 60 % RH, and 75 % RH in (a) longitudinal direction with an applied force of 10 µN and in (b) transverse direction with an applied force of 5 µN.*

To fit the data properly, the procedure previously described in the Materials and Methods section has been applied. To avoid too many fitting parameters and convergence problems, the values for the relaxation times were logarithmically evenly distributed for $n = 3$ relaxation modes over the experimental time of 240 s resulting in $\tau_1 = 1\ s, \tau_2 = 15\ s, \tau_3 = 240\ s$. Here, it should be emphasized that these relaxation times are a result of the experimental time scale (load schedule applied) and the continuum assumption. Consequently, additional relaxation times could be present in the material. They could be, for example, spatially localized on the different structural levels and are, therefore, not accessible with the AFM technique. However, since these relaxation times cannot be observed within the experimental limitations given, the restriction to the above three relaxation times is sufficient.

Representative experimental curves in longitudinal and transverse direction with the corresponding GM3 fit at 60 % RH are presented in Figure 5. To display the quality of the fit, 60 % RH has been chosen as a representative humidity stage. The GM3 for an experimental curve in longitudinal direction is presented in Figure 5a with a zoom-in at the first 10 s in Figure 5b. The same is illustrated for the transverse direction in Figures 5c and d.

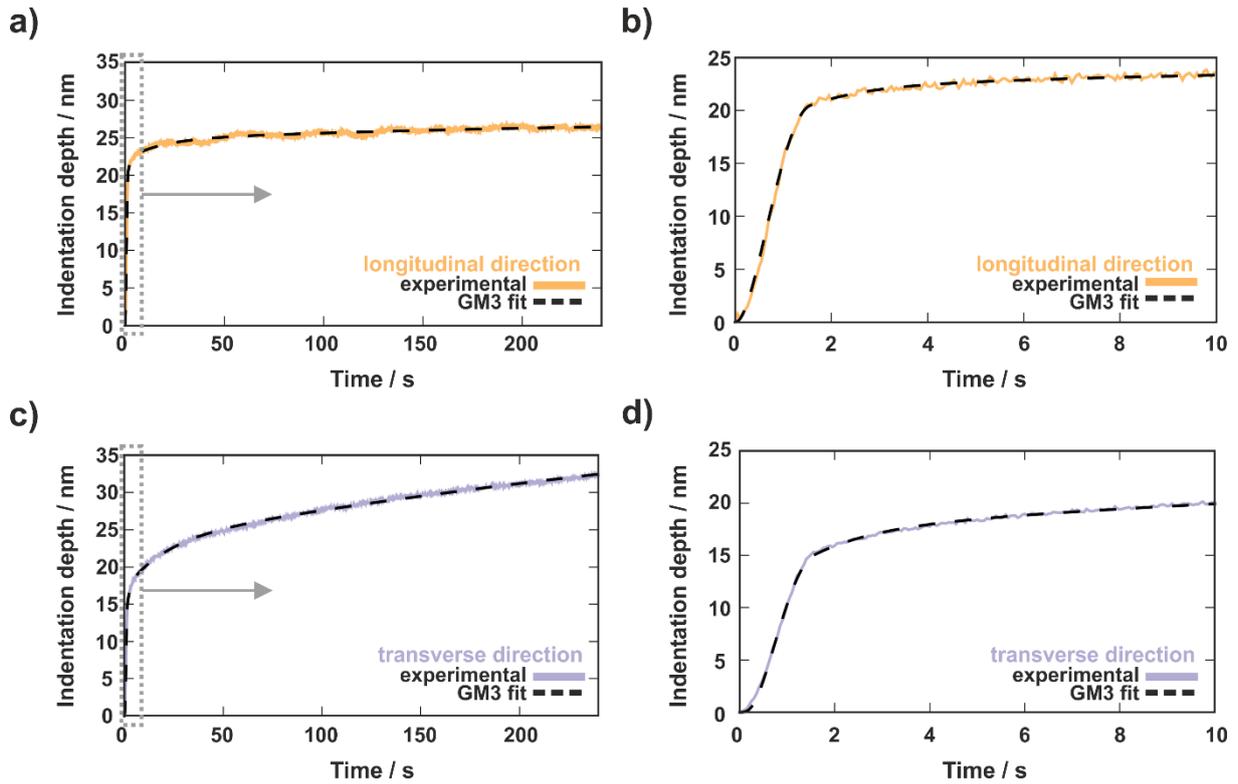

*Figure 5: Comparison of the fits of the GM3 model for the same representative experimental curve of a measurement position at 60 % RH in (a, b) longitudinal and (c, d) transverse direction. The experimental creep curves (orange and violet) are illustrated with the GM3 fit as black dashed lines. In (b, d), the first 10 s of the same experiments are presented at an expanded time scale.*

Results for the viscoelastic characterization in both directions and their dependence on relative humidity are presented in Figure 6. The presented results for all RH levels have been obtained with the GM3 model. The values in the diagrams are mean values obtained from 16 fibers in longitudinal direction and 6 fibers for the transverse direction. They are also presented in Table 1.

As can be seen in Figure 6a, the elastic parameters are described by $E_\infty$ and $E_0$. $E_\infty$ is the elastic modulus at infinitely slow loading, whereas $E_0$ is the elastic modulus at infinitely fast – instantaneous – loading. Both elastic moduli in longitudinal direction have higher values than in the transverse direction. Whereas $E_0$ is quite similar for both directions, the $E_\infty$ values show a larger difference between the directions up to 60 % RH. For the viscosities in Figure 6b, only slight differences between both directions can be found in Figure 6b. Overall, the viscosities are in the same range and stay quite constant over the whole RH range. Only at 75 % RH, $\eta_1$ and $\eta_3$ exhibit higher values for the longitudinal direction. Comparing only the values of $\eta_3$ in Figure 6b, one can observe a slightly different trend with RH. Whereas $\eta_3$ in transverse direction is decreasing with increasing RH, $\eta_3$ in longitudinal direction is increasing.

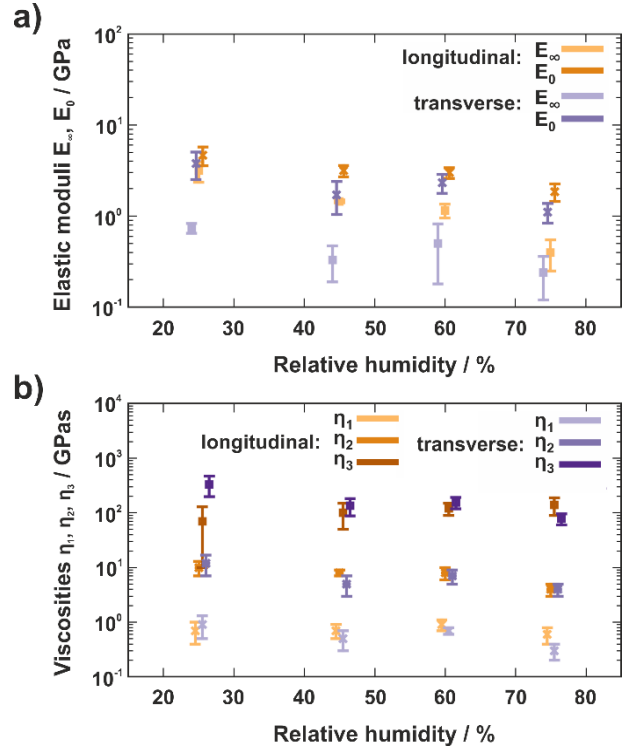

*Figure 6: Results of the viscoelastic characterization for the GM3 model with $\tau_1 = 1$ s, $\tau_2 = 15$ s, and $\tau_3 = 240$ s in longitudinal and transverse direction at different RH. (a) Elastic parameters $E_\infty$ and $E_0$, and (b) viscous parameters $\eta_1$, $\eta_2$, and $\eta_3$. The results are plotted as mean values ± confidence interval of 95 %.*

*Table 1: Results for the viscoelastic properties in longitudinal and transverse direction evaluated with the GM3 model with the fixed relaxation times $\tau_1 = 1$ s, $\tau_2 = 15$ s, and $\tau_3 = 240$ s. The values are given as mean ± confidence interval of 95 %. In longitudinal direction, results have been obtained from 16 individual fibers: 7 positions on 3 fibers (71 indents) for 25 % RH, 5 positions on 3 fibers (42 indents) for 45 % RH, 8 positions on 6 fibers (81 indents) for 60 % RH, and 7 positions on 4 fibers (54 indents) for 75 % RH. In transverse direction, results have been obtained from 6 individual fibers: 8 positions (61 indents) for 25 % RH, 14 positions (123 indents) for 45 % RH, 15 positions (117 indents) for 60 % RH, and 9 positions (64 indents) for 75 % RH.*

| RH | $E_\infty$ / GPa | $E_1$ / GPa | $E_2$ / GPa | $E_3$ / GPa | $E_0$ / GPa | $\eta_1$ / GPas | $\eta_2$ / GPas | $\eta_3$ / GPas |
|---|---|---|---|---|---|---|---|---|
| longitudinal | | | | | | | | |
| 25 % RH | 3.15 ± 0.80 | 0.70 ± 0.40 | 0.70 ± 0.20 | 0.30 ± 0.25 | 4.65 ± 1.10 | 0.7 ± 0.3 | 10 ± 3 | 70 ± 60 |
| 45 % RH | 1.45 ± 0.05 | 0.75 ± 0.20 | 0.50 ± 0.10 | 0.40 ± 0.20 | 3.15 ± 0.45 | 0.7 ± 0.2 | 8 ± 1 | 100 ± 50 |
| 60 % RH | 1.15 ± 0.20 | 0.85 ± 0.20 | 0.50 ± 0.10 | 0.50 ± 0.10 | 3.00 ± 0.40 | 0.9 ± 0.2 | 8 ± 2 | 120 ± 30 |
| 75 % RH | 0.40 ± 0.15 | 0.60 ± 0.20 | 0.30 ± 0.10 | 0.60 ± 0.20 | 1.85 ± 0.40 | 0.6 ± 0.2 | 4 ± 1 | 140 ± 50 |
| transverse | | | | | | | | |
| **25 % RH** | 0.74 ± 0.09 | 0.86 ± 0.36 | 0.78 ± 0.31 | 1.38 ± 0.56 | 3.77 ± 1.26 | 0.9 ± 0.4 | 12 ± 5 | 332 ± 135 |
| **45 % RH** | 0.33 ± 0.14 | 0.48 ± 0.21 | 0.34 ± 0.16 | 0.56 ± 0.19 | 1.72 ± 0.68 | 0.5 ± 0.2 | 5 ± 2 | 135 ± 47 |
| **60 % RH** | 0.50 ± 0.32 | 0.68 ± 0.14 | 0.49 ± 0.12 | 0.65 ± 0.15 | 2.33 ± 0.55 | 0.7 ± 0.1 | 7 ± 2 | 155 ± 37 |
| **75 % RH** | 0.24 ± 0.12 | 0.31 ± 0.09 | 0.23 ± 0.08 | 0.32 ± 0.08 | 1.11 ± 0.27 | 0.3 ± 0.1 | 4 ± 1 | 78 ± 18 |

Comparing the data in Table 1 for both directions, the absolute value of the infinite elastic modulus $E_\infty$ of the longitudinal direction is more than four times higher than in transverse direction. Furthermore, $E_\infty$ exhibits a higher decrease for the longitudinal direction from 25 % RH to 75 % RH. The value at 75 % RH is nearly a factor of eight lower than at 25 % RH. For the transverse direction, this decrease is not as high, the $E_\infty$ value is only about three times lower at 75 % RH compared to 25 % RH. Interestingly, for the instantaneous elastic modulus $E_0$, the difference in the absolute value of both directions is not as large as for $E_\infty$. At 25 % RH, the mean values of $E_0$ are within the confidence interval of both directions. Here, the $E_0$ value for the longitudinal direction has a lower decrease by a factor of 2.5 compared to 3.4 for the transverse direction. For the viscosities, also some differences between the directions have been found. The values of $\eta_1$ and $\eta_2$ are in a similar range between 25 % RH and 75 % RH for both directions, only $\eta_3$ is behaving differently. Whereas $\eta_1$ of the longitudinal direction is staying rather constant over the whole humidity range, $\eta_1$ of the transverse direction is decreasing by a factor of three. The value $\eta_2$ of is quite similar for both directions at all RH levels. However, $\eta_3$ exhibits an opposing trend. For the longitudinal direction, $\eta_3$ is increasing with increasing RH level by a factor of two, but for the transverse direction, the $\eta_1$ value is decreasing from 25 % RH to 75 % RH by a factor of 4.3.

In summary, the differences between longitudinal and transverse directions are still surprisingly small. As mentioned in the introduction, the longitudinal direction is expected to have a higher stiffness because of the microfibril reinforcement. Nevertheless, while performing mechanical testing on a scale that is comparable to the microfibrils' dimension which have a diameter of several tens of nm, the reinforcement can only be partially present resulting in a lower stiffness compared to uniaxial tension tests [47], that measures the complete structure. Interestingly, the instantaneous modulus $E_0$ does not render a large difference. This could be explained by the fact that in the contact initialization both, longitudinal and transversal direction have similar interaction with the stiff microfibrils. As the relaxation of the matrix progresses, the microfibrils can rearrange better in transverse direction than in longitudinal direction. Consequently, more parts of the matrix are loaded and the infinite modulus in transverse direction appears lower as in longitudinal direction.

Furthermore, it should be noted that the MFA of the investigated fibers is unknown and, therefore, its influence cannot be accounted for. There are optical techniques based on polarization [48] available to determine the MFA of single fibers, however, application to wood pulp fibers is not straightforward and requires substantial knowledge of the fiber structure.

In literature, numerous studies of conventional NI on the S2 layer of wood and wood pulp fibers [49–51] resulted in lower elastic moduli for the longitudinal direction than have been obtained with tensile testing. As a consequence of the fiber's anisotropy, the elastic modulus obtained with NI in longitudinal direction is not equal to the actual longitudinal elastic modulus. It is rather a mix of elastic stiffness components. Therefore, methods have been developed to extract the orthotropic elastic stiffness components from NI experiments on wood fibers [32,33]. A similar method might be also appropriate here, however, further experimental input like the viscoelastic shear properties is needed which is not available yet.

To obtain an additional visualization of the viscoelastic results, a spectral representation is applied to the data of the GM3 model for the longitudinal and the transverse direction. Since the relaxation times are $\tau_1 = 1\,s$, $\tau_2 = 15\,s$, and $\tau_3 = 240\,s$, the corresponding frequencies are the inverse values of the relaxation times, and a frequency range between 0.001 and 10 rad/s has been chosen for a complete visualization. The spectra of the storage modulus $E'(\omega)$, the loss modulus $E''(\omega)$, and the loss tangent $\tan(\delta(\omega))$ for the GM3 model are presented in Figure 7 for the longitudinal and transverse direction at the lowest (25 %) and the highest (75 %) RH level. They have been calculated by equations 3, 4, and 5 using the results from Table 1. In Figure 7a, it can be observed for both directions that the storage modulus is decreasing with increasing RH. Here, the values for the storage modulus in longitudinal direction are higher than in transverse direction. The inverse of the relaxation times $\tau_1$ and $\tau_3$ are the borders of the experimental window and indicate the lower and upper limits of the elastic moduli $E_\infty$ and $E_0$.

Figure 7b presents the results for the loss modulus. In longitudinal direction, the values show little change. With increasing RH, the peaks corresponding to $\tau_1 = 1\,s$ and $\tau_2 = 15\,s$ are slightly decreasing whereas the peak corresponding to $\tau_3 = 240\,s$ is slightly increasing. This indicates that the fast relaxation (short-term) behavior is moving towards the long-term behavior (described by $\tau_3$). The total viscoelastic softening, which is described by a decreasing loss modulus value, changes only slightly in longitudinal direction. For the transverse direction, however, a clear decrease of all three peaks with increasing RH is visible. The sharp peak at $\tau_3 = 240\,s$ which is present at 25 % RH decreases to the same level as the other two peaks at 75 % RH. Here, the long-term behavior is moving towards the short-term behavior. This is also the explanation for the opposing trend with increasing RH which has been found for $\eta_3$ (Figure 6b). Overall, the viscoelastic softening in transverse direction is higher than in longitudinal direction.

In Figure 7c the spectra of the loss tangent are presented. in the longitudinal direction. It exhibits an increase with increasing RH. At low RH, the values stay below 0.2, but the spectrum for 75 % RH shows a pronounced increase of the peak that is corresponding to $\tau_3 = 240\,s$. In transverse direction, the values for the loss tangent are larger and quite constant at all RH, only the peak corresponding to $\tau_3 = 240\,s$ exhibits a decrease with increasing RH. Taking a closer look, a similarity between the spectrum of 75 % RH in longitudinal direction and the transverse direction can be distinguished. This indicates that at 75 % RH the damping behavior of the longitudinal direction is quite similar to the behavior of the transverse direction at all RH.

In literature, the loss tangent for cellulosic materials has been investigated by dynamic mechanical analysis at different frequencies. Dynamic shear, bending, and compression tests have been carried out on fibers [52–54]. Here, a similar trend for the loss tangent at increasing humidity levels is observed. In general, the values for the loss tangent of wood, wood-based materials, and cork vary typically in the range between around 0.01 and 0.1 [55]. Pine cellulose fiber sheets at 20°C, which have been measured at a frequency of 100 Hz, have values for tan(δ) of 0.05 – 0.06 [56], whereas rayon fibers exhibit tan(δ) values between 0.04 – 0.05 for 10 Hz at room temperature [57]. Furthermore, recent Brillouin light scattering microspectroscopy (BLSM) measurements in the transverse direction of pulp and viscose fibers resulted in tan(δ) values of 0.04 in the GHz frequency range [58]. All these results fit well with the prediction of spectra at higher frequency in Figure 7c. Here, the spectra of both directions are well below a value of 0.1 at a frequency of 10 Hz.

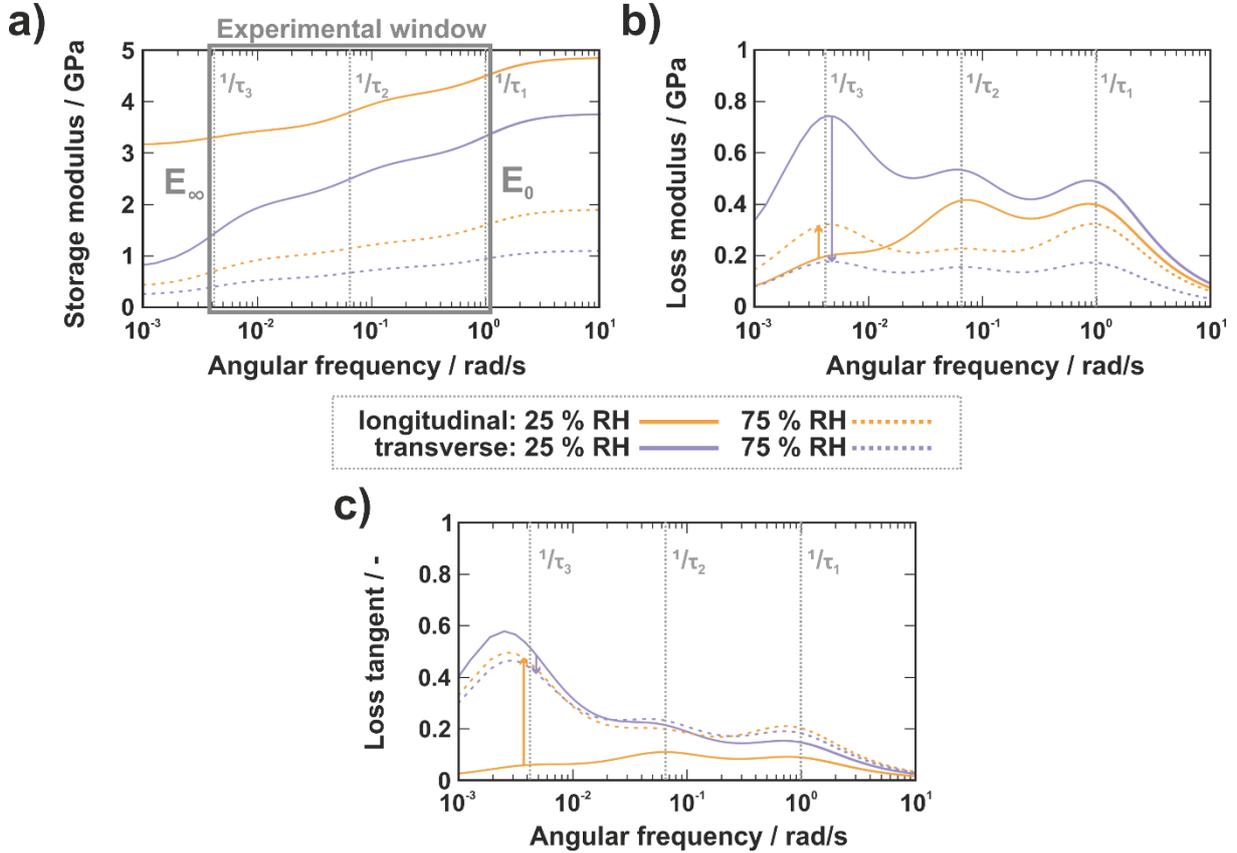

Figure 7: Spectral representations of the viscoelastic results in longitudinal and transverse direction for 25 % RH and 75 % RH (a) Storage modulus, (b) Loss modulus, and (c) loss tangent.

**Conclusions & Outlook**

In this work, a viscoelastic characterization of wood pulp fibers with an atomic force microscopy-based nanoindentation method (AFM-NI) has been presented and applied at different RH levels. The static creep measurements allowed the evaluation of the viscoelastic response of single wood pulp fibers in longitudinal and transverse directions at low frequencies. Here, a viscoelastic characterization in longitudinal and transverse directions measured with the same technique of a wood-based fiber is demonstrated for the first time.

Based on previous work [31,46], a generalized Maxwell model of order three (GM3) has been applied to the experimental creep data in both fiber directions, and the RH dependence of the elastic moduli $E_\infty$ and $E_0$ and the viscous parameters $\eta_1$, $\eta_2$, and $\eta_3$ has been studied. The results for both directions show a similar decreasing trend with increasing RH level.

The differences in viscoelastic behavior between longitudinal and transverse direction were found to be surprisingly small, the instantaneous elastic modulus $E_0$ is quite similar in both fiber direction at 25 % RH. At 75 % RH, $E_0$ is a factor of 2 higher for the longitudinal direction than for the transverse direction. A clearer difference is found for the infinite elastic modulus $E_\infty$ already at 25 % RH. Here, $E_\infty$ is more than four times higher in longitudinal direction than in transverse direction. For the viscosities, small differences between the directions have been found. The values of $\eta_1$ and $\eta_2$ are in a similar range between 25 % RH and 75 % RH for both directions, only $\eta_3$ is behaving in an opposing way for both directions.

The influence of the relative humidity increase is different for the elastic moduli. $E_\infty$ has a higher decrease from 25 % RH to 75 % RH for the longitudinal direction with a value that is nearly a factor of eight lower than at 25 % RH. In comparison, the $E_\infty$ value at 75 % RH is only about three times lower than at 25 % RH in transverse direction. However, the decrease of the instantaneous modulus $E_0$ is different. Here, the $E_0$ value for the longitudinal direction has a lower decrease by a factor of 2.5 compared to a factor of 3.4 for the transverse direction.

There are a few uncertainties that could have an influence on the results, e. g., the unknown microfibril angle (MFA) and the anisotropy of the material. Another point that needs consideration is that due to experimental limitations, for the transverse direction, the S1 layer was tested, whereas, in longitudinal direction, all measurements have been performed directly on the S2 layer.

With the application of a spectral representation of the results with storage modulus, loss modulus, and loss tangent, the experimental window between 0.004 Hz and 1 Hz can be better visualized. Overall, the values for the storage modulus in longitudinal direction are higher than in transverse direction, whereas the viscoelastic softening in longitudinal direction is lower than in transverse direction. The opposing trends found for the values of $\eta_3$ with increasing RH can be explained in terms of changes in short-term and long-term behavior. Furthermore, the damping behavior, which is described by the loss tangent, is much higher in transverse direction at low RH levels, however, it is similar to the longitudinal direction at 75 % RH. At higher frequencies (> 10 Hz), the loss tangent reaches values below 0.1 which fits well with literature values that have been recently obtained by BLSM.

For future measurements, the characterization of the MFA of single wood pulp fibers will be crucial. Furthermore, the development of a method that allows to extract the orthotropic elastic stiffness components from the presented viscoelastic AFM-based measurements would be very interesting for modeling approaches.

*Acknowledgments*

The financial support by the Austrian Federal Ministry for Digital and Economic Affairs and the National Foundation for Research Technology and Development is gratefully acknowledged. We also thank our industrial partners Mondi Group, Canon Production Printing, Kelheim Fibres GmbH, SIG Combibloc Group AG for fruitful discussions and their financial support. Special thanks to Angela Wolfbauer of the Institute of Bioproducts and Paper Technology, Graz University of Technology for sample preparation.

*Conflict of Interests*

The authors declare that they have no conflict of interest.

**Electronic Supplementary Information (ESI) for AFM-based low frequency viscoelastic characterization of wood pulp fibers at different relative humidity**

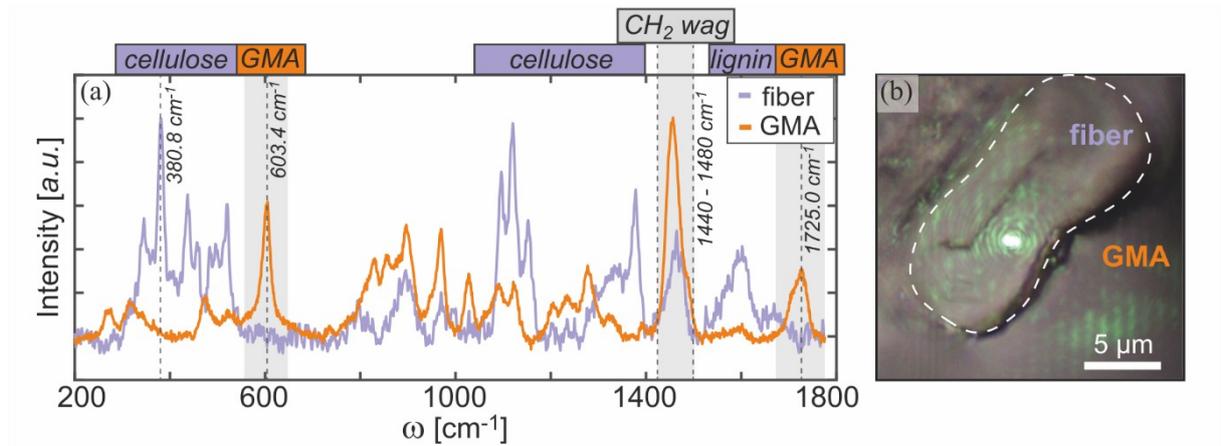

*Figure A1: (a) Raman spectra of a fiber and surrounding GMA. In the top, spectral ranges for the modes of cellulose, lignin, CH2 wagging, and GMA are marked. Two characteristic modes of GMA (at 603.4 cm-1 and 1725 cm-1) that were not observed in the fibers are highlighted. CH2 wagging mode (1440–1480 cm-1 range) and a characteristic cellulose mode at 380.8 cm-1 are also highlighted. (b) 20x20 µm2 optical microscopy image of the fiber. The dashed white line marks the fiber perimeter for clarity. The laser spot on the sample is visible and marks the spot from which the Raman spectra (a) of the fiber was measured. For the reported GMA spectra, the laser spot was positioned in the bottom-right corner of (b).*

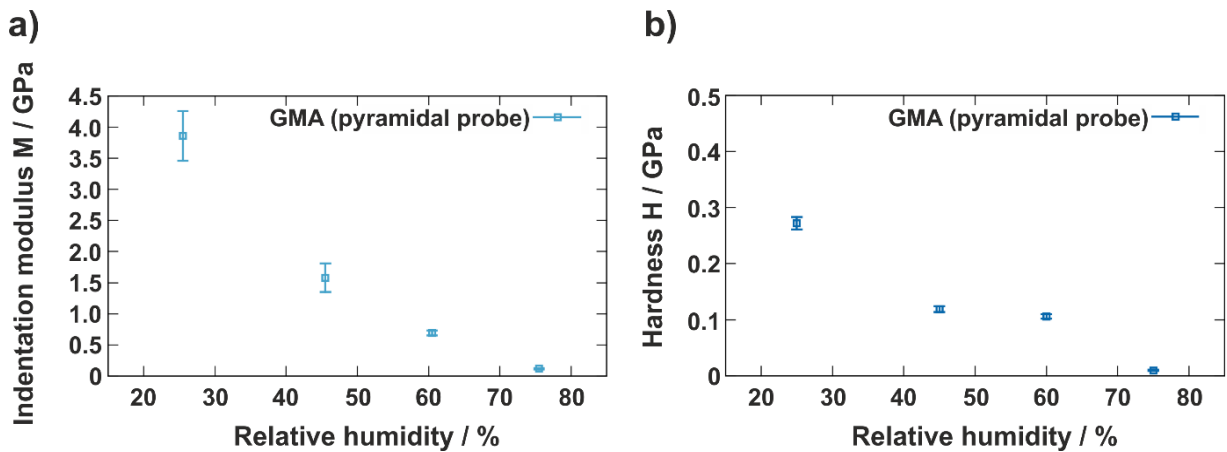

*Figure A2: (a) Indentation modulus $M$ and (b) hardness $H$ of glycol methacrylate (GMA) obtained with the pyramidal probe at different relative humidity levels.*